# Non-volatile memory based on PZT/FeGa thin film memtranstor

Jin-Cheng He(何金城)[1,2#], Jian Xing(邢健)[1#], Jian-Xin Shen(申见昕)[3], Dan Su(苏丹)[1,2], En-Ke Liu(刘恩克)[1], Shou-Guo Wang(王守国)[3,4*], and Young Sun(孙阳)[5†]

[1]*Beijing National Laboratory for Condensed Matter Physics, Institute of Physics, Chinese Academy of Sciences, Beijing 100190, China*
[2]*School of Physical Sciences, University of Chinese Academy of Sciences, Beijing 100190, China*
[3]*School of Materials Science and Engineering, Beijing Advanced Innovation Center for Materials Genome Engineering, University of Science and Technology Beijing, Beijing 100083, China*
[4]*School of Materials Science and Engineering, Anhui University, Hefei 230601, China*
[5]*Center of Quantum Materials and Devices, Chongqing University, Chongqing 401331, China*

The PZT/FeGa thin film memtranstor was prepared and the modulation of the magnetoelectric coefficient by external magnetic and electric fields was studied. The magnetoelectric coefficient of the PZT/FeGa memtranstor can be reversed by flipping the direction of magnetization of FeGa or ferroelectric polarization of PZT. Notably, the sign of the magnetoelectric coefficient can be switched repeatedly by reversing ferroelectric polarization of PZT when the external magnetic field remains constant. Moreover, the binary switching behavior can still be maintained under zero DC bias magnetic field. When the polarization direction remains stable, the magnetoelectric coefficient also does not change. This means that the magnetoelectric coefficient of PZT/FeGa is non-volatile. Furthermore, the retention and endurance characteristics of the PZT/FeGa thin film memtranstor have been investigated. These findings demonstrate the potential of the PZT/FeGa thin film memtranstor for non-volatile memory applications.

## 1. Introduction

With the continuous development of information technology, the need for a universal nonvolatile random-access memory (NVRAM) has been a topic of interest for several decades. Various physical effects have been utilized to develop memory devices, such as magnetic random-access memory (MRAM),[1-3] resistive switching random-access memory (RRAM),[4,5] phase change random-access memory (PCRAM),[6-7] and ferroelectric random-access memory (FeRAM).[8-10] However, these devices face challenges that limit their industrial application. Recently, a novel device based on the magnetoelectric coupling effects, called memtranstor, has emerged as a promising NVRAM.[11-13] Memtranstor exhibits superior performance, not only due to its electric field (*E*) write and magnetic field (*H*) read work style, but also because of its fast write speed, low energy cost, and non-destructive readout. The magnetoelectric coefficient α$_E$, which can be either positive or negative, depending on the magnetoelectric coupling mechanism and the status of magnetic moment (*M*) and electric polarization (*P*), is used to store digital data, instead of *M* or *P* themselves. Therefore, memtranstor represents a new approach for developing NVRAM devices with high performance and potential for widespread use in practical applications.

[#] These authors contributed equally to this work and should be considered co-first authors
[*] Corresponding author. E-mail: sgwang@ustb.edu.cn
[†] Corresponding author. E-mail: youngsun@cqu.edu.cn

In our previous study, memtranstors made of bulk multiferroic heterostructures such as PMN-PT/Terfenol-D,[13-15] Ni/PMN-PT/Ni,[16,17] Metglas/PMN-PT[18] and FeGa/PMN-PT/FeGa[19] have been prepared to demonstrate the functions as a NVRAM,[13,14,18] Boolean logic gate[15,16] and artificial synaptic device.[17,19] However, due to dimension issues, these bulk memtranstors have limited practical applications. In this point, the development of thin film devices is of great importance. In this paper, we report on a PZT/FeGa thin film memtranstor, with both FeGa and PZT having a thickness of 200 nm, and demonstrate its nonvolatile random-access memory capabilities.

## 2. Experiments

As show in Fig. 1(b), the memtranstor was prepared by depositing FeGa onto the surface of a PZT thin film using magnetic sputtering to form PZT/FeGa heterostructures. The PZT thin film was prepared as the ferroelectric component by spin coating onto a Pt/Sapphire substrate. This method is widely used to prepare oxide thin film.[20-22] The thicknesses of both the PZT and FeGa films were 200 nm. Although Terfenol-D and Metglas have higher magnetostrictive effect, FeGa is more suitable due to its simple composition and significant magnetostrictive effect at zero magnetic field.[23,24] FeGa also acted as an electrode due to its conductivity, while Pt served as another electrode. PZT is well-known for its significant piezoelectric effect, and the preparation process of PZT film has been extensively studied and is becoming increasingly mature.[25,26] The magnetoelectric (ME) coupling of the PZT/FeGa heterostructure comes from the interfacial strain between the magnetic and ferroelectric layers. A conventional dynamic technique was used to measure the voltage of magnetoelectric coupling (*V*$_{ME}$).[27,28] As shown in Fig. 1(a), a small AC magnetic field (*H*$_{AC}$, ~2 Oe) which is generated by a solenoid with frequency of 10 kHz and a DC magnetic field (*H*$_{DC}$) were applied in-plane to the PZT/FeGa heterostructure. The magnetic moments (*M*) of FeGa were aligned by *H*$_{DC}$, and the ferroelectric polarization (*P*) of PZT was tuned by an external electric field (*E*). FeGa undergoes a periodic distortion under the action of *H*$_{AC}$ due to the magnetostrictive effect. This periodic distortion is transferred to PZT through the interfacial strain, resulting in a periodic change of *P*, which produced magnetoelectric coupling voltage (*V*$_{ME}$). The induced *V*$_{ME}$ signal was picked up using a Stanford Research SR830 lock-in amplifier. To switch the ferroelectric polarization of the PZT thin film, ±5 V voltage pulses with a 10 ms width were applied out-of-plane to the memtranstor. All measurements were performed at room temperature.

## 3. Results and Discussion

Figure 2 displays the $V_{ME}$ characteristics of the PZT/FeGa memtranstor as a function of $H_{DC}$ under positive (red line) and negative (black line) ferroelectric polarizations after applying +5 V and -5 V voltage pulses, respectively. The reduction in thickness of PZT film leads to a decrease in coercive field to 5 V, rendering the PZT/FeGa memtranstor compatible with most electric circuits. The sign of $V_{ME}$ can be changed by switching the *P* or *M*. The $V_{ME}$ was measured after applying a +5 V pulse to pre-pole the ferroelectric polarization direction upward (the red curve). The $V_{ME}$ is very small at high $H_{DC}$ due to the saturation of FeGa magnetostriction. As the $H_{DC}$ is reduced to 0, the $V_{ME}$ gradually decreases to a minimum of approximately -0.8 µV. When the magnetic field sweeps to the negative direction, the $V_{ME}$ increase rapidly to a maximum of approximately 1 µV, and then returns to zero as $H_{DC}$ increases toward the negative field. When the $H_{DC}$ sweeps from negative to positive, $V_{ME}$ increases to a maximum of approximately 0.8 µV and then decreases to a minimum of approximately -1 µV as the filed reaches positive. Finally, $V_{ME}$ gradually decreases to about zero when the $H_{DC}$ above 1 kOe. For the downward ferroelectric polarization direction (the black curve), the behavior of $V_{ME}$ is completely opposite. Due to the saturation of FeGa magnetostriction, $V_{ME}$ remains very small at high $H_{DC}$. With $H_{DC}$ decreasing, the $V_{ME}$ gradually increase to a maximum of approximately 0.8 µV. When the magnetic field sweeps to the negative direction, the $V_{ME}$ rapidly decreases to a minimum of approximately -1 µV, and returns to zero as $H_{DC}$ falls below -1 kOe. When the $H_{DC}$ sweeps from negative to positive, $V_{ME}$ decreases to a minimum of approximately -0.8 µV and then increases to a maximum of approximately 1 µV as the filed reaches positive. Finally, $V_{ME}$ gradually decrease to about zero when the $H_{DC}$ above 1 kOe. The opposite behavior of $V_{ME}$ is the key feature that makes the memtranstor a memory device. It is notable that $V_{ME}$ is near maximum at zero $H_{DC}$ due to the self-bias effect,[29-31] which makes PZT/FeGa memtranstor more practical.

The ME coupling in memtranstors arises from the interfacial strain between the magnetic and ferroelectric layers. When a magnetic field is applied, it induces strain in the magnetic layer via the magnetostriction effect, which is then transferred to the piezoelectric layer. The strain in the piezoelectric layer results in an electric displacement through the piezoelectric effect, leading to the generation of a magnetoelectric coupling voltage.

In order to verify the ability of the PZT/FeGa thin film memtranstor as a memory device, the binary states have been repeatedly switched for many times. As shown in Fig. 3(a), a -5 V pulse with a width of 10 ms was applied to the memtranstor, and the $V_{ME}$ was measured for 100 seconds. Then, a +5 V pulse with the same width was applied and the $V_{ME}$ was measured for another 100 seconds. This process was repeated for 10 cycles. And all measurements were conducted under zero $H_{DC}$. The $V_{ME}$ was repeatedly switched between the maximum and minimum states after the application of the voltage pulse. Moreover, the $V_{ME}$ remained stable until another voltage pulse was applied, which demonstrated the nonvolatile memory feature of the PZT/FeGa thin film memtranstor.

We carried out a further study on the non-volatile memory performance of

PZT/FeGa thin film memtranstor. In practical applications, device retention and endurance are critical parameters. Fig. 4(a) shows the the retention of the PZT/FeGa thin film memtranstor, where the $V_{ME}$ of the two states was measured over a period of 6000 seconds, respectively. The value of $V_{ME}$ remained stable during the measuring process. Meanwhile, the binary states of the PZT/FeGa thin film memtranstor were repeatedly switched to examine the endurance character. After 100 switching cycles, the values of the two states remained stable and could be clearly distinguished. Based on the measurement results, the PZT/FeGa thin film memtranstor exhibits outstanding retention and endurance characteristics.

It is necessary to emphasize that storing information using magnetoelectric coefficient of PZT/FeGa thin film memtranstor offers several notable advantages. Firstly, the simple sandwich structure of the PZT/FeGa thin film memtranstor makes it easy to prepare and allows for reduced device size. Secondly, the writing operation is simple, fast, and energy-efficient, similar to FeRAM and RRAM, and achieved by applying voltage pulses to both ends of the electrodes. Third, the read process of the memtranstor avoids the switching of direction of *P* like FeRAM and the complexity of *M* change read in MRAM. Moreover, the magnitude of the magnetoelectric coefficient is independent of the area of the memtranstor, avoiding the limitation on the minimum size of the memory cell in FeRAM and enabling a significant increase in the theoretical storage density of the memtranstor.

4. **Conclusion and perspectives**

In summary, we have successfully prepared and investigated the non-volatile random memory performance of PZT/FeGa thin film memtranstor. The magnetoelectric coupling voltage ($V_{ME}$) of the device can be tuned by the external magnetic field and the direction of ferroelectric polarization of PZT. Moreover, the $V_{ME}$ can be repeatedly switched by changing the direction of ferroelectric polarization of PZT through external voltage pluses under zero magnetic field. By reducing the thickness of the device, the operating voltage has been lowered to 5 V, making it suitable for most circuits. Additionally, the PZT/FeGa thin film memtranstor demonstrates excellent retention and endurance characteristics, which further support its potential as a future non-volatile memory device.

**Acknowledgment**

This work was supported by the National Key Research and Development Program of China (Grant Nos. 2020YFF01014703, 2021YFA1400303) and National Nature Science Foundation of China (Grant No. 52101281).

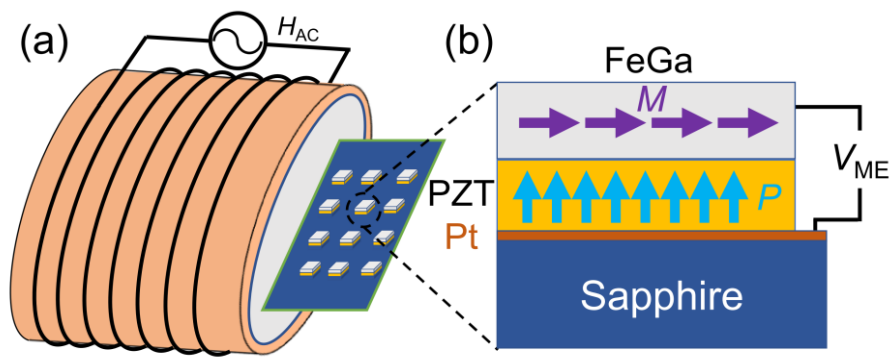

Figure1. Structure and principle of PZT/FeGa thin film memtranstor. (a) The array of memtranstors is placed into a readout solenoid, which generates an AC magnetic field ($H_{AC}$). (b) The structure of the PZT/FeGa thin film memtranstor with in-plane magnetization in the FeGa layer and out-of-plane polarization in the PZT layer. FeGa and Pt play the role of electrodes.

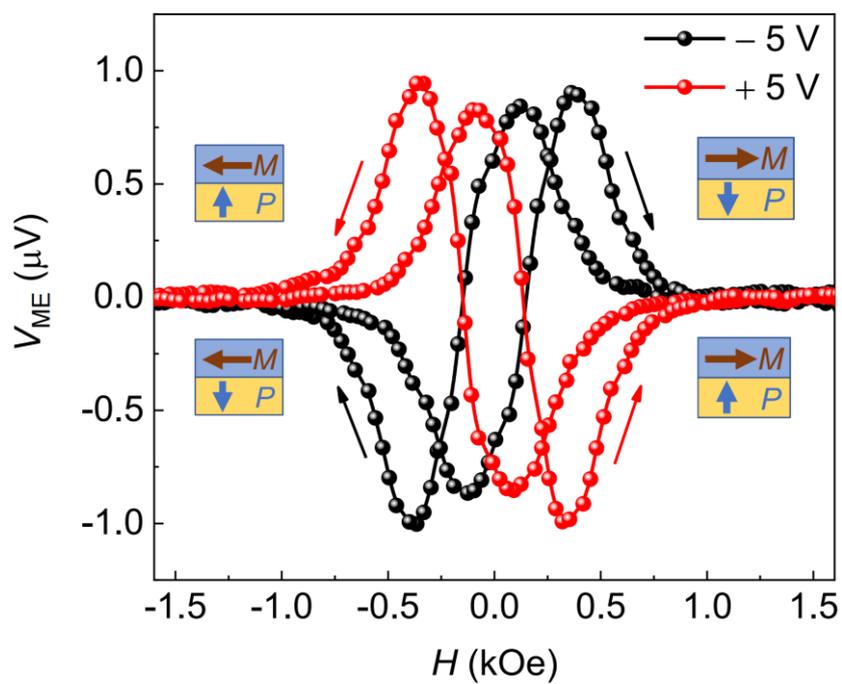

Figure 2. The ME voltage of PZT/FeGa thin film memtranstor as a function of DC bias magnetic field after positive (red) and negative (black) poling. The $H_{AC}$ is about 2 Oe with a frequency of 10 kHz. The sign of the ME voltage can be controlled by selecting the direction of polarization as well as the direction of magnetic moment.

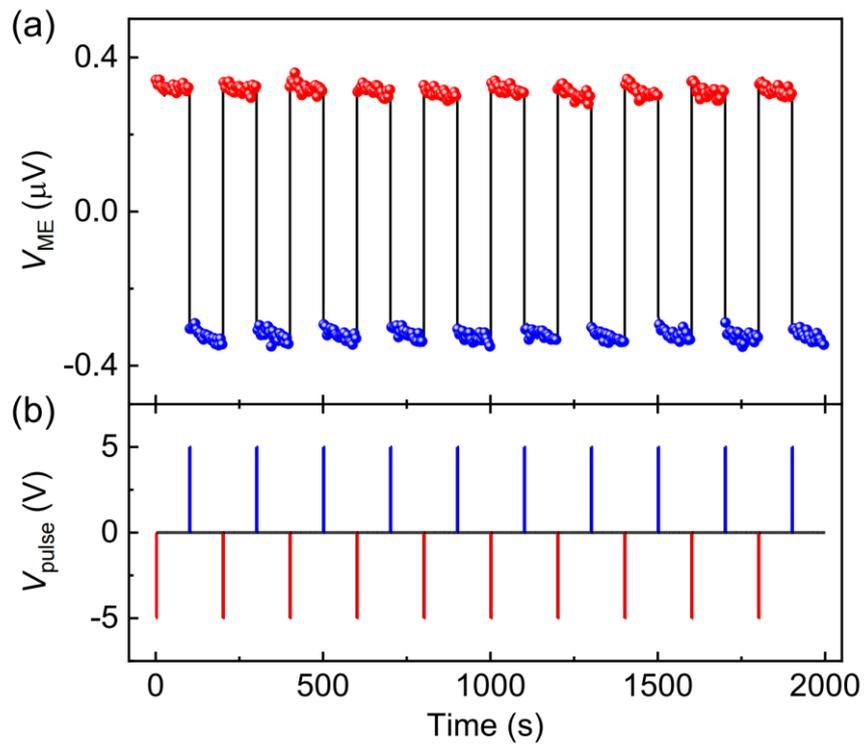

Figure 3. Two-level non-volatile memory based on the PZT/FeGa thin film memtranstor. (a) Repeatable switch of the ME voltage between positive and negative under zero bias magnetic field. (b) The sequence of voltage pulses. After applying a ±5 V pulse with a 10 ms width to reverse electric polarization, the ME voltage was measured for 100 s.

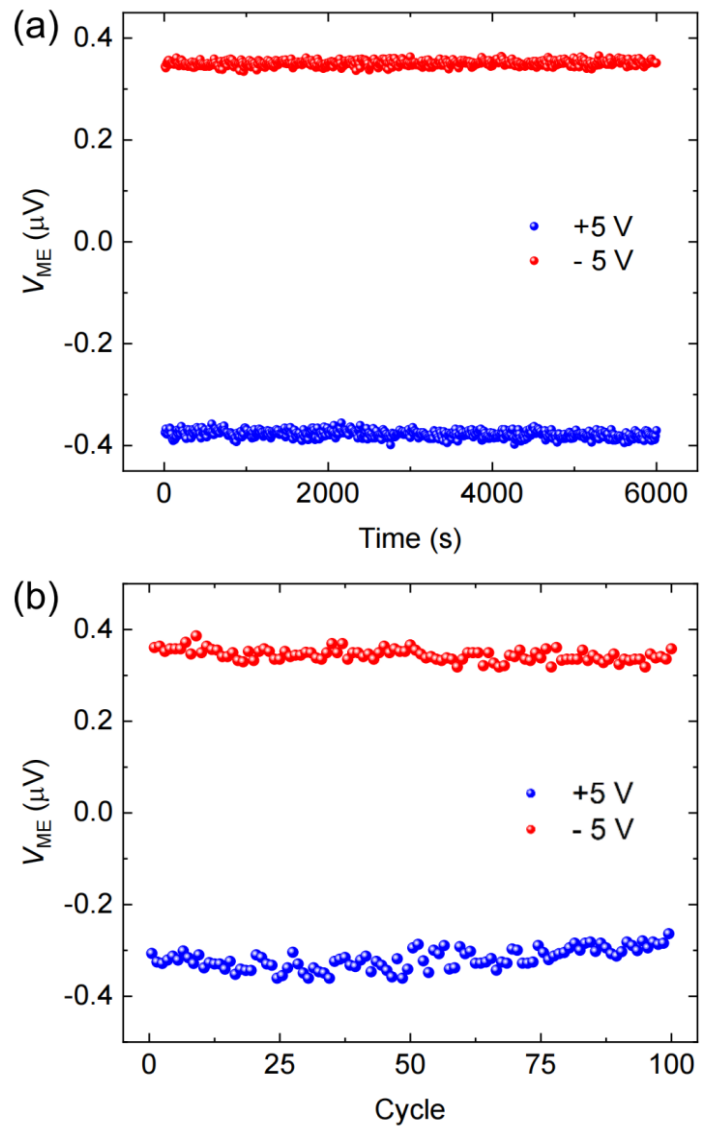

Figure 4. The retention and endurance characteristics of PZT/FeGa thin film memtranstor. (a) The $V_{ME}$ of two states was measured over 6000 seconds, respectively. (b) The binary states of PZT/FeGa thin film memtranstor were repeatly switched over 100 cycles.